\documentclass[aps,prb,preprint]{revtex4}

\begin{document}

\title{Properties of ferroelectric nanodots \\ embedded in a polarizable medium: Atomistic simulations}

\author{S. Prosandeev$^{1,2}$ and L. Bellaiche$^1$}

 \affiliation{$^1$Physics Department, University of Arkansas, Fayetteville, AR 72701\\
$^2$Institute of Physics, Rostov State University, Rostov on Don,
Russia 344090}

\date{\today}

\begin{abstract}

An atomistic approach is used to investigate finite-temperature
properties of ferroelectric nanodots that are embedded in a
polarizable medium. Different phases are predicted, depending on
the ferroelectric strengths of the material constituting the dot and
of the system forming the medium. In particular, novel states,
exhibiting a coexistence between two kinds of order parameters or
possessing a peculiar order between dipole vortices of adjacent
dots, are discovered. We also discuss the origins of these phases,
e.g. depolarizing fields and medium-driven interactions between dots.
\end{abstract}
\maketitle

Ferroelectric nanostructures (FENs) are of current high interest,
because of their technological promise in leading toward
miniaturized devices \cite{Scott}, and because of their potential in
resulting in new phenomena. For instance, is was recently predicted
that {\it isolated} nanodots of ferroelectrics can have a vortex
structure for their dipoles below a critical temperature
\cite{Naumov,Ponomareva,HuaxiangPRL}. Such vortex does not create
any polarization but rather generates a macroscopic toroidal moment,
that involves the cross product between the ${\bf r}_i$~ vectors
locating the $i$ unit cells and their local electrical dipoles ${\bf
p}_i$ -- i.e., it is defined as ${\bf G}=\frac{1}{2N}\sum_i{{\bf
r}_i \times {\bf p}_i}$, where the sum runs over the $N$ unit cells
of the nanodot. Being able to switch the direction of ${\bf G}$
opens exciting opportunities for  nanomemory devices
\cite{Naumov,Dubovik2}.

Interestingly, the possibility for the dipoles to form a vortex
structure has been omitted in the previous studies (see, e.g., Refs.
\cite{effmed1,effmed2,effmed3} that used an effective medium
approximation) aimed in determining the properties of ferroelectric
or paraelectric nanoparticles that are {\it embedded} in a
polarizable medium. As a result,  original features may have been
overlooked in these composite systems. For instance, one may wonder
if novel states occur in such two-component materials. In
particular, can dipole vortices in dots immersed in a polarizable
medium coexist with a spontaneous polarization (unlike in isolated
dots \cite{Naumov,Ponomareva})? Similarly,  how do toroidal moments
of neighboring dots organize themselves? (e.g, are they lying along
parallel or antiparallel directions?). It would also be worthwhile
to determine the dependency of these novel states (if any) on the
ferroelectrics strengths of the materials forming the dot and
medium, and to reveal their governing mechanisms. The aim of this
Letter is to address the issues mentioned above, by performing
atomistic simulations.

Our atomistic scheme is based on the construction of an
effective Hamiltonian, with the total energy $E$ being written as a sum of
two main terms,
\begin{eqnarray}
   E (\{ { \bf u_{\it i}} \}, \{ \eta_{\it H} \}, \{  \eta_{\it I} \},  \{ \sigma_{\it j} \} )&&
      =
   E_{\rm ave} (\{ { \bf u_{\it i}} \}, \{ \eta_{\it H} \},\{ \eta_{\it I}\})
   +~E_{\rm loc} (\{ { \bf u_{\it i}} \},\{ \eta_{\it I}  \}
  ),
\end{eqnarray}
where $E_{\rm ave}$ -- as in Refs. \cite{PRLPZT,APLPSN} -- is the
total energy associated with the hypothetical {\it simple}
A$<B>$O$_{3}$ system resulting from the use of the virtual crystal
approximation (VCA) \cite{LaurentDavid3} to mimic A(B$'$,B$''$)O$_3$
compounds;  $E_{\rm loc}$  gathers alloying terms going beyond the
VCA approximation; ${\bf u_{\it i}}$ is the local soft mode
(directly related to the electrical dipole) in unit cell $i$; $\{
\eta_{\it H} \}$ and $\{  \eta_{\it I} \}$ are the {\it homogeneous}
and {\it  inhomogeneous} strain tensor, respectively
\cite{ZhongDavid} ; $\{ \sigma_{{\it j}}\}$ characterizes the atomic
configuration, that  is,  $\sigma_j = 1$ (resp. $-$1) if there is a
B$'$ (resp. B$''$) cation at the B-lattice site $j$ of the
A(B$'$,B$''$)O$_3$ materials. For $E_{\rm ave}$, we use the
analytical expression proposed in Ref.~\cite{ZhongDavid} for {\it
simple} ABO$_{3}$ systems.  For  $E_{\rm loc}$, we use the following
expression:
\begin{eqnarray}
 E_{\rm loc} (\{ { \bf u_{\it i}} \},\{ \eta_{\it I} \},\{ \sigma_{\it j} \} ) =
  \sum_{ij} [Q_{\it j,i} \sigma_{\it j} ~ { \bf e_{\it ji}} \cdot { \bf u_{\it i}}~+~
  R_{\it j,i} \sigma_{\it j} ~ { \bf f_{\it ji}} \cdot { \bf v_{\it i} }] +
  \sum_{i} \Delta \kappa (\sigma_{\it i}) ~ u_{\it i}^{2}
\end{eqnarray}
where the sums over $i$ and $j$ run over unit cells and
mixed sublattice sites, respectively.
$\{ { \bf v_{\it i}} \}$ are dimensionless local displacements
which are related to the inhomogeneous strain variables inside each
cell \cite{ZhongDavid}. $Q_{{\it
j,i}}(\sigma_{\it j})$ and $R_{{\it j,i}}(\sigma_{\it j})$ characterize the strengths of
the alloying-induced intersite interactions. $\bf{e_{\it ji}}$ is a unit vector
joining the site $j$ to the center of the soft-mode vector $\bf{u_{\it
i}}$, and $\bf{f_{\it ji}}$ is a unit vector joining the site $j$ to the
origin of the displacement $\bf{v_{\it i}}$.  Practically,
 we included contributions up to the third neighbors for
$Q_{\it j,i}(\sigma_{\it j})$, and over the first-neighbor shell for
$R_{\it j,i}(\sigma_{\it j})$.  Note that, for systems made of
AB$''$O$_3$ dots embedded in a AB$'$O$_3$ medium, these
contributions only play a role near the dots' surfaces -- because of
the analytical expression of the first two  terms of
Eq.~(2). The last term of Eq.~(2), which involves the $\Delta \kappa
(\sigma_{\it i})$ parameters, characterizes the onsite contribution
of alloying. It is an original contribution since it
was not included in the alloy effective Hamiltonians  of
Refs.~\cite{PRLPZT,APLPSN}. We incorporate such term here
 -- which is consistent with first-principles results of Ref. \cite{Bungaro} --
because it provides an easy way to artificially adjust  the
``ferroelectric strength'' of AB$'$O$_3$ and AB$''$O$_3$ simple
systems, by playing with the $\Delta \kappa (\sigma_{\it i}=+1)$ and
$\Delta \kappa (\sigma_{\it i}=-1)$ parameters. For instance, a
large negative (respectively, positive) $\Delta \kappa(\sigma_{\it
i}=-1)$ leads to a strong ferroelectric instability (respectively,
no ferroelectric instability) of the pure  AB$''$O$_3$ material. In
the following, we will denote $\Delta \kappa$ as the difference
between the two alloying-onsite parameters, i.e.  $\Delta
\kappa=\Delta \kappa(\sigma_{\it i}=+1) - \Delta \kappa(\sigma_{\it
i}=-1)$.
At the exceptions of the adjustable $\Delta \kappa (\sigma_{\it i}=+1)$ and $\Delta \kappa (\sigma_{\it i}=-1)$ variables, all the
 parameters of our
toy model described by Eqs.~(1)-(2) are those derived for the Pb(Zr$_{0.5}$Ti$_{0.5}$)O$_3$ solid solution from first-principles
calculations \cite{PRLPZT}. [Such parameters
yield a tetragonal ferroelectric ground-state with a polarization pointing along a $<001>$ direction
 and a Curie temperature $\simeq$ 1000\,K, in the Pb(Zr$_{0.5}$Ti$_{0.5}$)O$_3$ bulk].
As a result, we numerically found that  $\Delta \kappa
(\sigma_{\it i}=+1)$ or $\Delta \kappa (\sigma_{\it i}=-1)$ $\simeq$ +0.0094 a.u. is the highest
value to have a ferroelectric ground-state in pure  AB$'$O$_3$  or AB$''$O$_3$ material, respectively.

The total energy of Eq.~(1) is used in Monte-Carlo simulations to
obtain the local mode vectors, the toroidal moment of polarization and
the homogeneous strain tensor $\{ \eta_{\it H} \}$ at different
temperatures.  Typically, 2$\times$10$^{4}$ Monte-Carlo sweeps are
first performed to equilibrate the system, and then
8$\times$10$^{4}$ sweeps are used to get various statistical
averages. The temperature is decreased in small steps down to 1\,K.
We also heat up the investigated systems by small temperature steps from
1\,K, to look for thermal hysteresis. We also computed the dielectric
susceptibility, as well as the  toroidal susceptibility (which
characterizes the response of the toroidal moment to the curl of the
electric field \cite{Gorbatsevich}), as in Ref. \cite{Alberto1}.

Figure 1 shows the temperature-versus-$\Delta \kappa$ phase diagram
of a $16\times 16\times 16$ periodic supercell (20480 atoms, 64\AA
~lateral size) containing an AB$''$O$_3$ cubic dot of 48\AA~of lateral
dimension embedded in a host matrix made of pure AB$'$O$_3$. Practically, for {\it positive} $\Delta \kappa$,
$\Delta \kappa (\sigma_{\it i}=-1)$ is set
to zero while $\Delta \kappa (\sigma_{\it i}=+1)$ is allowed
to vary. These cases thus correspond to
a specific ferroelectric dot  immersed in
different media that are all ferroelectrically-harder than the dot and that can
either be ferroelectric (small positive $\Delta \kappa$) or
paraelectric (larger positive $\Delta \kappa$). The
reverse situation applies for the case of {\it negative} $\Delta
\kappa$: $\Delta \kappa (\sigma_{\it i}=-1)$ is  varied while
$\Delta \kappa (\sigma_{\it i}=+1)$ vanishes,  implying that we
mimic either ferroelectric dots with smaller ferroelectricity
strength than the medium (for small negative $\Delta \kappa$) or
paraelectric dots (for larger negative $\Delta \kappa$) inserted in a matrix
made from a specific ferroelectric material.

Figure 1 reveals the existence of six different phases, for which
the associated dipole patterns are displayed in insets [Phase
boundaries were practically determined by identifying the peak or
abrupt jumps of the susceptibilites altogether with the appearance
of the spontaneous polarization or toroidal moment. Note that, in
principle, further study could reveal some of these features to be
associated with continuous changes in configuration rather than true
phase boundaries {\it per se}. However, some preliminary results
obtained by us on the basis of the finite-size scaling analysis
\cite{Binder2} do confirm the existence of true phase transitions.]
Two of these phases are expected based on previous knowledge of
ferroelectrics: the paraelectric, PE, state at high temperature, and
the ferroelectric, FE1, phase occurring at intermediate and low
temperature when $\Delta \kappa$ is small in magnitude. In this FE1
state, both the dot and medium develop homogeneous parallel dipoles
-- with these dipoles being larger (resp., smaller) in the dot than
in the host when $\Delta \kappa$ is positive (resp., negative), as
consistent with our definition of $\Delta \kappa$ (i.e., the
difference between the onsite parameters of the medium and the dot).

Four phases of Fig.~1 can be considered as novel structures. They
are denoted as FT, FE1+FT, FE2 and FE3, respectively. The FT phase,
which occurs for the largest positive $\Delta \kappa$ values,
exhibits a finite toroidal moment. The appearance of this state in
this region of the phase diagram results from the fighting of the
dipoles in the dot against the large enough depolarizing field
\cite{Naumov,Ponomareva,HuaxiangPRL} (which arises from the
significant non-similarity between the ``more ferroelectric'' dot
and ```less ferroelectric'' host material). The present discovery of
this FT phase is, in fact, consistent with the previous finding that
{\it isolated} ferroelectric dots surrounded by vacuum exhibit a
vortex structure for their dipoles below a critical temperature
\cite{Naumov,Ponomareva,HuaxiangPRL}, because one can think of
vacuum as a medium having an infinite positive value for $\Delta
\kappa (\sigma_{\it i}=+1)$. However and unlike in the vacuum, the
dipoles of the host matrix in the FT phase become slightly polarized
by the nearby dipoles located inside the dots and near the surfaces,
when this host matrix is still rather ``soft'' (see corresponding
inset of Fig.1). For such cases, the medium thus also generates a
toroidal moment that is parallel to, but of lower magnitude than,
the one solely associated with the dot.

Moreover, the FE1+FT phase  appearing in the phase diagram at small
temperature and within a narrow range of positive $\Delta \kappa$ is
rather remarkable  because it displays an unusual cohabitation
between the toroidal moment and the spontaneous polarization. In
this phase, the medium generates an electric field inside the dot,
that (as in some magnetic nanodots under an external magnetic field
\cite{Guslienko} or in some ferroelectric dots subject to electric
fields \cite{Sergey1}) leads to the shift of the vortex structure
with respect to the center of the dot (see the corresponding inset
in Fig. 1) and thus activates a polarization. Note, too, that this
FE1+FT phase can be considered as a low-temperature bridging
structure between the FE1 and FT phases since we numerically found
that the FE1--to--FE1+FT and FE1+FT--to--FT transitions are
second-order in character, unlike the FT--to-FE1 phase transition
that displays all the expected features of a first-order phase
transition (e.g., the toroidal moment suddenly disappears at this
boundary in favor of a \emph{finite} value of the polarization, and
the FT--to--FE1 boundary line has a rather large thermal hysteresis:
it is typically increased by $\simeq$ 100 K with respect to the one
displayed in Fig. 1 when heating, rather than cooling).

The last two phases, FE2 and FE3, appearing in Fig.~1 are both
ferroelectric and occur at intermediate and low temperatures,
respectively, for the largest negative values of $\Delta\kappa$. In
other words, these two states correspond to  cases for which the
dot, unlike the medium, is made of a material that desires to be
paraelectric. As a result, the FE2 phase exhibits a spontaneous
polarization that originates from the alignment along a $<001>$
direction of dipoles belonging to some {\it specific regions of the
medium}. More specifically, these regions belong to the $\{001\}$
planes that contain the polarization direction and that do not
possess any site belonging to the dot (see, e.g, the corresponding
inset of Fig. 1 for FE2 showing the homogeneity of dipoles in the
top and bottom planes, while the parts of the medium located at the
right and left sides of the dot do {\it not} display any homogeneity
for their dipoles). The reason behind such unique arrangement, in
which the largest dipoles in the medium intentionally avoid to point
towards the dots,  is once again the minimization of the
depolarizing energy. When decreasing the temperature, the FE2 phase
transforms into the FE3 state, that is associated with a
polarization that has now two non-vanishing components along two
different $<001>$ directions. More precisely, the corresponding
inset of Fig. 1  reveals that, in the FE3 phase, the top and bottom
parts of the medium exhibit dipoles that are similar in direction
than those in the FE2 phase, while dipoles located in the medium at
the right and left sides of the dot have dipoles aligned along a
{\it perpendicular} direction. Such dipole arrangement arises, once
again,  from a minimization of the depolarizing energy.
Interestingly, the FE2 and FE3 phases bear resemblance to some
states that were experimentally found recently in artificially
constructed magnets \cite{antidots}. Moreover, note that  we
numerically found that another solution (but of slightly higher
energy than that of the FE3 phase) is possible at low temperature
for large negative $\Delta\kappa$: it consists of the top and bottom
parts of the medium having dipoles homogeneously aligned a specific
$<001>$ directions (as in the FE2 phase) while the dipoles of the
medium located at the right and left sides of the dot form a vortex
structure.

We also numerically checked
that all the phases of Fig.~1 still occur when varying the size of
the dot or the size of the whole supercell. Moreover,  out of these
six states, the FT and FE1+FT phases are the soles structures that
refine themselves when allowing {\it several}  AB$''$O$_3$ dots to
be present in a supercell possessing a  AB$'$O$_3$  medium.
More precisely, Figs 2 show that, for large enough positive
$\Delta\kappa$,  neighboring dots have vortices rotating in an {\it
opposite} fashion. In other words, the FT (respectively, the FE1+FT) phase of Fig.1 should in fact become the AFT
 (respectively, FE1+AFT) structure of Fig. 2a (respectively, Fig. 2b) when dots are close enough from each other (as in dots' arrays).
 Such novel {\it antiferrotoric} phases, unlike the FT and FT+FE1 states,  allow the dipoles located {\it between} two adjacent dots to all point along  similar directions, which
 minimizes the corresponding domain wall energy at the cost of the electrostatic interaction between neighboring vortices (we numerically found that  the direct electrostatic interaction between two different toroidal moments should align these moments along
 the same direction -- that is, it should lead to a ferrotoric, rather than antiferrotoric, order).  Furthermore, Fig.~2b reveals that the centers of the vortices in the
FE1+AFT phase form a rather unusual lattice (see, e.g., that the
centers of the vortices have different y-locations for dots that are
adjacent along the x-axis).
This unique geometry originates from the desire of the whole system
to maximize the number of its dipoles (in the dots) lying along the
polarization direction, considering the underlying antiferrotoric
order.

In summary, varying the polarizabilities of the embedded
dots and of the medium opens a new way for engineering dipole
patterns. The
fundamental reasons behind such engineering are depolarizing-fields
effects and medium-driven interactions between dots. Interestingly,
our findings may result in promising applications using negative refraction materials \cite{Vendik}, nanomemory devices, etc....
They should also be relevant to, and thus lead to revisiting the
analysis of,  samples (i) made of composites \cite{BST-SiO2};  (ii)
in which segregation takes places (e.g.,
Ba(Zr,Ti)O$_3$ for intermediate compositions \cite{Farhi});  (iii) in which chemically-ordered
nanoregions form inside a disordered matrix (e.g., relaxors
such as Pb(Mg,Nb)O$_3$ \cite{Davies}).  One can also easily imagine that further considering multi-component (e.g., three-component)
systems can lead to even
richer phase diagrams. Furthermore, designing some specific {\it lattice} of ferroelectric dots
(inside some medium)  may lead to other exotic states --
as similar to the recent finding of Ref. \cite{Ice} for nanoscale ferromagnetic islands.

We thank Igor Kornev for useful discussions. This work is mostly
supported by DOE grant DE-FG02-05ER46188. We also acknowledge
support from ONR grant N00014-04-1-0413 and NSF grant DMR-0404335.
Some computations were made possible thanks to the MRI Grant 0421099
from NSF. S.P. also thanks RFBR grants 01-02-629\&05-02-90568\_HHC.

\newpage

\section*{CAPTIONS}
Fig. 1.
Temperature-versus-$\Delta \kappa$ phase diagram
of a  $12\times 12\times 12$ AB$''$O$_3$  dot embedded
in a AB$'$O$_3$ medium within a $16\times 16\times 16$ periodic supercell.The positive $\Delta \kappa$ part of
this diagram corresponds to a soft ferroelectric dot immersed in a medium that is ferroelectrically harder than the dot and that has
a decreasing ferroelectric instability as $\Delta \kappa$ increases. The negative $\Delta \kappa$ part of this diagram corresponds
to a dot (having a ferroelectric instability that is weaker than those of the medium and that decreases, and then vanishes,
as $\Delta \kappa$ increases in magnitude) embedded in a ferroelectrically-soft medium.
The lines with symbols represent the phases' boundaries.
The insets show a (001) cross-section of the dipole configuration in the different phases. Specifically, these insets correspond to
atomistic calculations with the following ($\Delta \kappa$, temperature) combination:
(-0.0212 a.u.,  1K),    (-0.0212 a.u., 500K),    (0.0062  a.u.,  1K),   (0.0087 a.u.,  1 K)  and (0.0112 a.u.,  1K)
for  the FE3, FE2, FE1,  FE1+FT  and FT phases, respectively. The dot surfaces are indicated via thick continuous lines in these insets.
The x- and y-axes are chosen along the pseudo-cubic [100] and [010] directions, respectively.

\vspace{2mm}

Fig. 2. (001) cross-sections of the dipole configuration in the AFT and FE1+AFT phases in panels a and b, respectively,
for four $12\times 12\times 12$ AB$''$O$_3$  dots embedded in a AB$'$O$_3$ medium within a $32\times 32\times 32$ periodic supercell.
Such cross-sections specifically correspond to atomistic calculations for which the ($\Delta \kappa$, temperature) combination is
(0.0112 a.u.,  1K) and  (0.0087 a.u.,   1K) for the AFT and FE1+AFT states, respectively. The dots
surfaces are represented by  thick continuous lines. The x- and y-axes are chosen along the pseudo-cubic [100] and [010] directions, respectively.

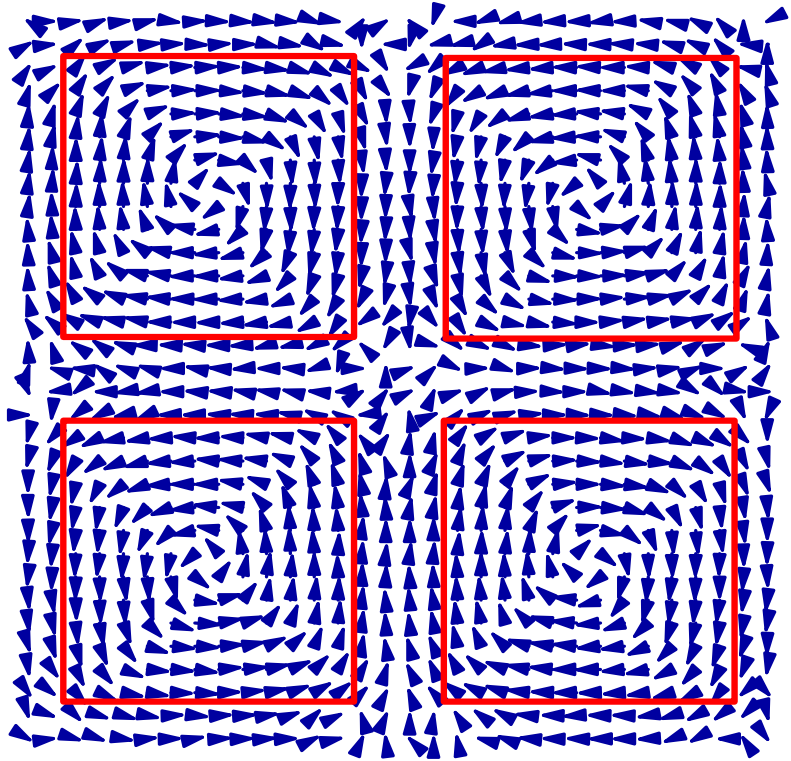
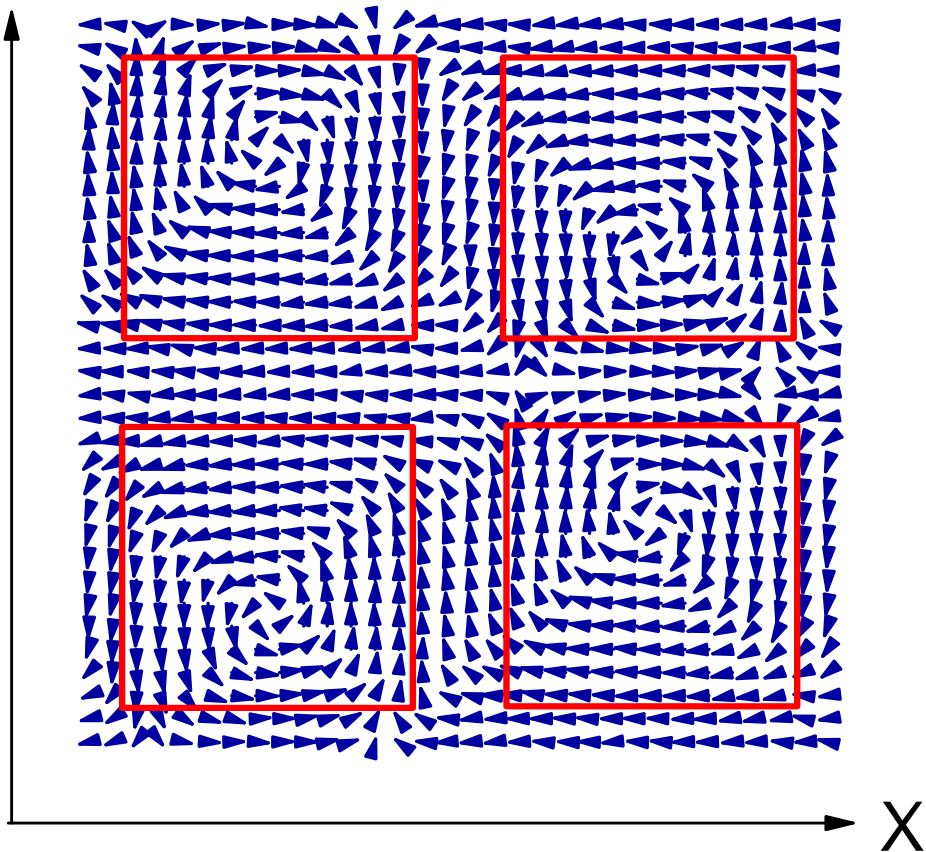
\end{document}